\documentstyle[12pt,aaspp4]{article}
\begin{document}

\title{A semianalytical model for the observational properties
of the dominant carbon species at different metallicities}

\author{Alberto D. Bolatto, James M. Jackson and James G. Ingalls}
\affil{Department of Astronomy, Boston University, Boston MA 02215}
\authoraddr{Boston University Dept. of Astronomy\\725 Commonwealth Ave.\\
Boston, MA 02215}

\begin{abstract}
We present a simple, semianalytical approach to compute the line emission
of C$^+$, C$^0$ and CO in Photodissociation Regions of varying metallicity
that depends on very few parameters and naturally incorporates the 
clumpiness of the interstellar medium into the calculations. 
We use it to study the dependence of the 
${\rm I_{[C II]}/I_{CO}}$ and ${\rm I_{[C I]}/I_{CO}}$ line ratios on the 
metallicity. We show how to modify it to include the effects of
density and radiation field, and we have compared it with observational
data. We find that the model explains the observed trend of enhanced 
${\rm [C~II]/CO}$ line ratio with decreasing metallicity as the natural result
of the augmented fraction of photodissociated gas in a clump. We show
that enhanced ${\rm [C~II]/CO}$ ratios can be produced by lowering the 
upper limit of the clump size distribution, as may happen in 30 Doradus.
We also find that the available data favors a ${\rm [C~I]/CO}$ intensity 
ratio essentially independent of metallicity, albeit the paucity of 
observations does not exclude other possibilities. 
This is difficult to understand if most of 
the C$^0$ is produced by UV photons as is the case with C$^+$.
Finally, we study the prediction of the model for the trend of
the $X_{\rm CO}$ factor with metallicity. Comparison with previous 
observational studies yields good agreement.
\end{abstract}


\section{Introduction}
\subsection{Photodissociation regions and the metallicity of the ISM}

A photodissociation region (PDR or photon dominated region) is that portion 
of the interstellar medium where ultraviolet (UV) radiation dominates 
the physical and chemical processes.  The first theoretical studies of PDRs
(e.g., Tielens and Hollenbach 1985a\markcite{TH85a}; Wolfire, Tielens, \& 
Hollenbach 1990\markcite{WTH90}) 
considered the strong UV fields in the vicinity
of OB stars.  These models and others have shown that UV
photons not energetic enough to ionize hydrogen
($h\nu > 13.6$ eV) escape the H~II region and interact with the surrounding
molecular gas. These soft UV photons 
photodissociate molecules and ionize atoms with ionization potentials 
less than 13.6 eV, such as carbon, silicon, and sulfur,
creating a transition zone between the ionized and molecular gas. 

Theoretical PDR models (e.g., Tielens and 
Hollenbach 1985a\markcite{TH85a}) predict that the
primary coolants of PDRs are far-infrared (FIR) fine-structure atomic lines,
particularly those of [O~I] and [C~II].  Hence, these lines are
expected to be quite
bright near regions of massive star formation.  This expectation
is borne out by observation: the [C~II] 158 $\mu$m line alone 
accounts for 0.1 to 1\% of the total far-infrared luminosity in spiral galaxies
(Stacey et al. 1991\markcite{SG91}), corresponding to a luminosity of up to
$\sim 10^9$ L$_\odot$.  The [C II] line is by far the most luminous 
spectral line in galaxies.
Because the FIR lines are so bright, they are often used to infer the 
physical conditions in gas near recently formed massive stars.

Wolfire et al. (1990)\markcite{WTH90} suggest that the most 
important observational
diagnostics for PDRs are: (1) CO, which provides an estimate of the molecular
gas mass, (2) [C~II] and [O~I] FIR fine structure line emission,
which provide information on the densities and temperatures, and
(3) FIR continuum emission, which, in combination with (1) and (2) provide
the incident UV flux, the stellar luminosities, and the dust opacities.
Recently, [C~I] submillimeter line emission has also been shown
to be an important diagnostic of the PDR/molecular cloud interface
(e.g., Plume, Jaffe, \& Keene 1994\markcite{PJK94}; 
Keene et al. 1993\markcite{K93}), especially 
in translucent clouds with $A_V \sim 1 - 2$ (Ingalls, Bania, \& Jackson 1994
\markcite{IBJ94}; Ingalls et al. 1997\markcite{I97}).

Although there has been considerable progress in the study of
PDRs, it is still difficult to separate the influences of 
various physical parameters such as UV field strength and metallicity 
on the structure and composition of the interstellar medium. 
In this paper we examine the effects of metallicity on the
structure of PDRs and the intensity of their diagnostic lines. 

Both theory and observation suggest that metallicity
profoundly alters the properties and structure of
the photodissociation regions (e.g. Israel 1988\markcite{IS88}; 
Maloney 1990\markcite{MA90}; 
Madden et al. 1997\markcite{MA97}).
Several factors contribute to the differences between high and low
metallicity systems.
Because low metallicity systems have fewer heavy elements,
the gas phase chemistry is altered, resulting in
different equilibrium abundances of the molecular species
(e.g., van Dishoeck \& Black 1988\markcite{VDB88}; 
Lequeux et al. 1994\markcite{LQ94}).
Grain surface chemistry is also altered because a deficiency of heavy elements
(especially Si, C and S) lowers the dust-to-gas ratio.
Finally, since there is less dust shielding, UV photons penetrate
more deeply into molecular clouds.
Molecular hydrogen is largely unaffected by this enhanced
UV radiation because of the strong H$_2$ self-shielding in the Werner
and Lyman bands and the mutual shielding between 
coincident lines of H and H$_2$
(Abgrall et al. 1992\markcite{AB92}). However, molecules which
are not so strongly self-shielding, such as CO,
are destroyed via photodissocation
everywhere except in the most opaque clumps
(Maloney \& Black 1988\markcite{MB88}, Elmegreen 1989\markcite{E89}).
Thus, a low-metallicity system contains large regions where hydrogen
remains molecular but the usual tracers of molecular gas like CO
are absent.
Accordingly, molecular gas in low-metallicity
systems is not well traced by CO.
 
The paucity of CO emission from molecular gas in low-metallicity
systems is confirmed by observations of the ISM in
metal-poor dwarf galaxies.  Dwarf galaxies
systematically display very low levels of CO emission
(e.g., Tacconi \& Young 1987\markcite{TY87}).
CO observations of the Large Magellanic Cloud (LMC, Cohen et al.
1988\markcite{CO88}; Israel et al. 1993\markcite{IS93}) 
and the Small Magellanic Cloud 
(SMC, Rubio, Lequeux, \& Boulanger 1993\markcite{RLB93}) show that
the ratio between the CO luminosity and the virial
mass is much smaller in LMC and SMC clouds than that in Galactic clouds.
Furthermore, a comparison of outer Galaxy clouds with those
in the SMC (where the metallicity is
1/6 relative to the Orion Nebula) shows that CO is less luminous 
by a factor of $10-20$ over large spatial scales ($\sim 200$ pc)
(Rubio et al. 1993\markcite{RLB93}).
Wilson (1995)\markcite{WI95} and Arimoto, S\={o}fue, \& 
Tsujimoto (1996)\markcite{AR96} have examined the value of the
$X_{\rm CO}$ factor, the ratio between the CO luminosity and the hydrogen
column density, in galaxies spanning a large range
of metallicities.  They find a strong dependence of $X_{\rm CO}$ 
on metallicity: for a given molecular gas mass,
lower metallicity systems emit less CO $J=1\rightarrow0$.
These results all suggest that CO is fainter in low-metallicity
clouds due to enhanced photodissociation.
 
Observations of the [C II] 158 $\mu$m line also
provide strong evidence that photodissocation regions
are greatly enhanced in dwarf galaxies.
Ionized carbon is the end product of the photodissociation of CO.
Recent KAO observations of the [C~II] 158 $\mu$m transition in the LMC 
(Poglitsch et al. 1995\markcite{PO95}, Israel
et al. 1996\markcite{IS96}) and in IC~10 (Madden et al. 1997\markcite{MA97}) 
show that the [C~II] emission is extended and extremely bright.
In fact, the ratio of [C~II]/FIR continuum emission in the LMC is about $1\%$,
much higher than in the Galaxy and most galactic nuclei.
Furthermore, the [C~II]/CO luminosity ratios for H II regions
in the LMC are larger than those in the Milky Way
by a factor of $\sim 5$ for N160 and a factor
of $\sim 50$ for 30 Doradus (Israel et al. 1996\markcite{IS96}).
Similar results are found for the northern dwarf galaxy IC 10 
(Madden et al. 1997\markcite{MA97}).
The dense ISM in dwarf galaxies seems to be a vast
photodissociation region, where CO is present only in well shielded
clumps.

To better understand the transition between ionized, neutral atomic, and
molecular forms of carbon in low-metallicity dwarf galaxies, we have 
begun a project to study their submillimeter [C~I] lines.
Neutral carbon emission was detected from a low-metallicity
system for the first time using the Antarctic Submillimeter
Telescope and Remote Observatory (AST/RO), a new 1.7 m diameter
telescope recently deployed at the South Pole 
(Stark et al. 1998\markcite{SB98}).
The [C~I] line has been measured measured towards two positions in the LMC
(Stark et al. 1997\markcite{SB97}):
N~159, the star forming region near one of the brightest
CO $J=1\rightarrow0$ peaks in the LMC, and 30 Doradus, 
the brightest H~II region in the local Universe .
We have also detected [C~I] emission from the dwarf galaxy IC~10 using the 
JCMT at Hawaii (Bolatto et al. 1998\markcite{BO98}). 

These new data prompt us to model the intensities of
the various forms of carbon in molecular clouds of varying 
metallicity. 

\subsection{Modelling low metallicity PDRs}

There are several models of photodissociation regions
that treat the chemistry, radiative transfer and energy balance of molecular
clouds in a detailed, self consistent and highly sophisticated way 
(e.g.: Tielens \& Hollenbach 1985a\markcite{TH85a}; 
van Dishoeck \& Black 1988\markcite{VDB88}; 
Le Bourlot et al. 1993\markcite{LB93}; 
Sternberg \& Dalgarno 1995\markcite{SD95}; 
Spaans 1996\markcite{SP96}; Pak et al. 1998\markcite{PA98}). 
Many of them perform the calculations in a plane-parallel, 
uniform density geometry. However, molecular clouds are clumpy objects, 
highly permeable to the UV radiation that drives most of the 
chemistry in the PDR (e.g., Stutzki et al. 1988\markcite{ST88}). 
Applying the
results of plane-parallel calculations in the context of clumpy PDRs 
(e.g., Poglitsch et al. 1995\markcite{PO95}; Madden et al. 1997\markcite{MA97})
necessitates the additional step of assuming different 
beam filling fractions for different species; otherwise the observed 
line ratios of the different carbon 
species cannot be explained. This problem is even more acute for low 
metallicity systems, where the UV field permeates the clouds even more
pervasively, due to the higher gas to dust ratios and the smaller
dust shielding. 
Spaans and coworkers (e.g., Spaans 1996\markcite{SP96}; 
Spaans \& van Dishoeck 1997\markcite{SvD97}; Spaans \& 
Neufeld 1997\markcite{SN97}) have made numerical 2-D and 3-D Monte Carlo 
models of PDRs that incorporate the effects of clumpiness in a very 
complete way, albeit only for particular realizations of the 
distribution of clumps within the cloud. 

The purpose of the model presented here is to bridge the gap 
between the plane-parallel, uniform slab calculations and 
the numerical models of clumpy PDRs.
This model naturally incorporates the effects of
different beam filling fractions for the different carbon species, and
the predicted line ratios depend only on a few free parameters
that can be directly linked to observations. 
We do not intend to compute accurately the chemistry, radiative transfer
and energy balance. Instead, we use simple fits to the
chemistry calculations of previous plane-parallel models to include 
the effects of density and radiation fields in our calculations, and we 
rely on observations to determine the values of our model parameters.

In \S\ref{assumptions} we will state some of our 
initial simplifying assumptions,
in \S\ref{basic} we will develop the basic mathematics of
the model and compute the column densities of the carbon species, 
in \S\ref{lrats} we will compute the line intensity ratios,
in \S\ref{parval} we will discuss the values for the model parameters,
in \S\ref{densandrad} we will discuss how these values are affected by
changes in the gas density and radiation field, in \S\ref{comparison}
we will compare our model results with the available observations,
and in \S\ref{xcosection} we will discuss the relationship between
this model and the observed CO intensity to H$_2$ column density 
conversion factor.

\section{The model}
\label{assumptions}
We assume that the gas phase of the ISM is constituted by uniform density,
unresolved, spherical clumps of radius $R$ immersed in an isotropic UV field. 
The assumption of an 
isotropic UV field is justified in view of the large UV back-scattering 
coefficient for dust grains (e.g., Savage \& Mathis 1979\markcite{SM79}).
We consider a distribution of clump radii in the telescope beam of the 
form $dN/dR=R^{-\beta}$. 
This type of power law for the distribution of clump sizes
has been observed in molecular clouds with exponent $\beta\approx3.3$
(Stutzki \& G\"usten 1990\markcite{SG90}; Elmegreen \& 
Falgarone 1996\markcite{EF96}). 
To simplify the mathematics of the model we will take this 
exponent to be $\beta=3$. Changes of order 10\% in $\beta$ affect 
minimally the model results. The model implicitly considers unresolved clumps;
our resolution element is greater than the maximum clump size. The
observed intensities then result from the combined contributions of the 
ensemble of clumps within the telescope beam.
We use the same average density $n$ for all the clumps. 
Studies of the mass and size distributions of clumps in Milky Way molecular
clouds (Leisawitz 1990\markcite{LW90}; Elmegreen \& Falgarone 
1996\markcite{EF96} and references therein)
reveal that the clump masses follow a power law $M\propto R\ ^{3.1}$, i.e., the
clump volume density $n$ appears to be roughly constant. Thus, 
we are not incorporating a density distribution or a size-density 
relationship (Larson 1981\markcite{LA81}; Myers 1983\markcite{MY83}). 
Throughout this work we assume that the dust-to-gas ratio is proportional 
to the metallicity of the interstellar medium.
 
According to PDR theory, a typical clump has three well defined regions: 
an outermost zone where C$^+$ is the dominant form of carbon, a core where 
almost all the carbon is locked in CO molecules, and a transition region in
between where most of the carbon is in neutral atomic form (C$^0$). 
We model each clump as consisting of three concentric zones with 
sharp boundaries, where all the carbon is in the dominant form of the species
corresponding to each zone.

\section{The column density of the carbon species}
\label{basic}\subsection{Model A}

Let the physical extent of the C$^0$ region be $h$. We assume 
that the following scaling for $h$ with metallicity holds

\begin{equation}
h={h_0 \over Z}
\label{cscaling}
\end{equation}

\noindent where $Z$ is the metallicity of the gas 
relative to the Milky Way (i.e.,
$Z=1$ for Milky way objects) and $h_0$ thus is the width of the C$^0$ region
in clumps of $Z=1$. We can justify this scaling law by the following argument:
the abundance of C$^0$ is determined by the balance between production of
C$^0$ via photodissociation of CO, and destruction of C$^0$ via 
photoionization into C$^+$. The rates for both photoprocesses are 
dominated by dust shielding of the UV photons, while other processes
that do not scale with the metallicity (such as CO cross-shielding with H$_2$) 
play only a secondary role. Observationally, we
know that the dust-to-gas ratio scales approximately as the metallicity 
(e.g., Koornneef 1982\markcite{KO82}; Bouchet et al. 1985\markcite{BO85}), 
which is not surprising since heavy elements are needed to 
create grains. Therefore the extent of the region over which both 
photoprocesses balance will be inversely proportional to the 
metallicity. In other words, the extent of the C$^0$ region is constant
in visual extinction (A$_{\rm v}$), but since the length scale corresponding
to A$_{\rm v}$ scales as $1/Z$ so does $h$. Note that the CO self-shielding,
which depends on the CO abundance and ultimately on the C content of the
gas, scales in the same manner as the dust shielding.

Similar considerations lead us to write the scaling law for the extent of 
the C$^+$ region ($h_{+}$) as

\begin{equation}
h_{+}={\alpha h_0 \over Z}
\label{cpscaling}
\end{equation}

\noindent where $\alpha$ is an assumed constant ratio between 
$h_{+}$ and $h$. From chemical modelling (e.g. Hollenbach, Takahashi, 
\& Tielens 1991\markcite{HTT91}) we surmise that $\alpha\sim1-2$.

\begin{figure}
\plotone{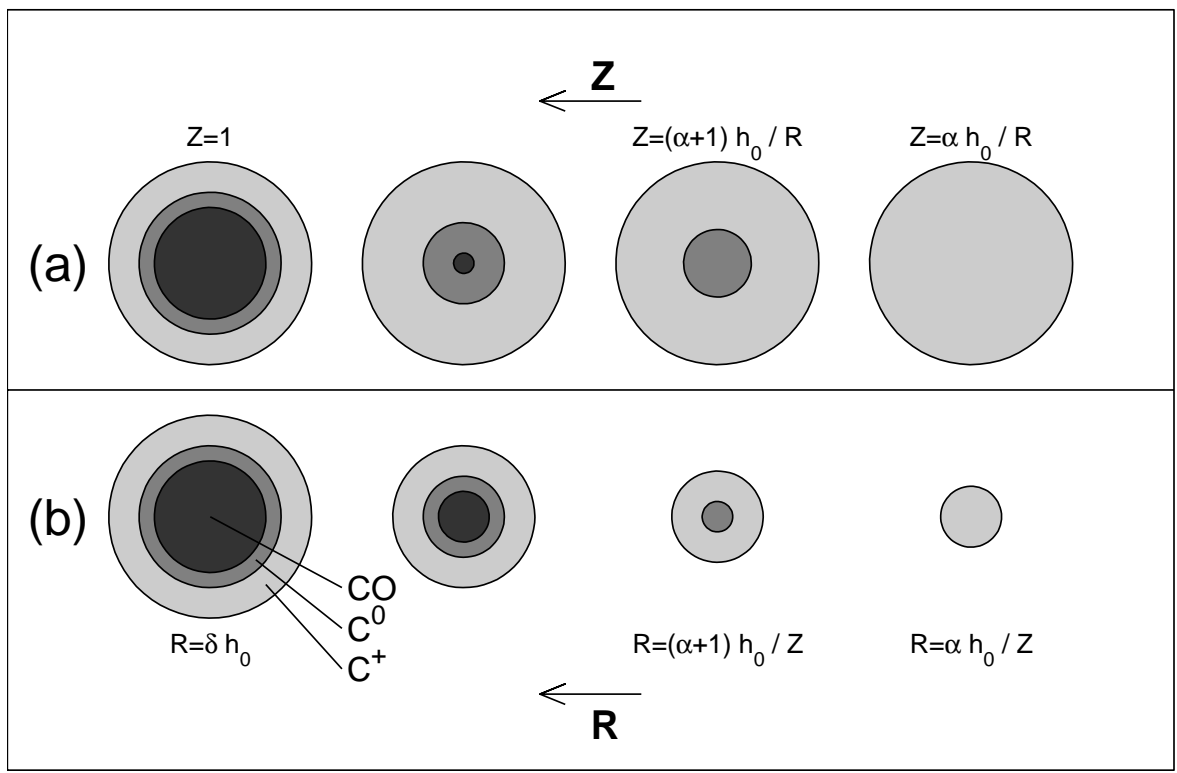}
\caption{{\em (a) Top Row.} A sequence of clumps with identical radius and 
decreasing metallicity according to {model A}. 
When $Z={(\alpha+1) h_0/R}$, the CO is completely 
photodissociated and the clump consists only of C$^0$ and C$^+$. The next
stage happens at $Z=\alpha h_0/R$, when all the neutral carbon is photoionized 
into C$^+$. At even lower metallicities all the gas consists only of C$^+$.
{\em (b) Bottom Row.} A sequence of clumps with constant metallicity and 
decreasing radii in {model A}. Similar sequences where the limiting cases
occur at slightly different metallicities can be created using {model B}.}
\label{clumpfig}
\end{figure}

All the material that is not photoionized (C$^+$) or photodissociated 
(C$^0$) is in molecular form (CO), and given the radius $R$ of a clump it is 
easy to compute the size of the CO region as $h_{\rm CO}=R-h-h_+$. 
Figure \ref{clumpfig}{\em a}
illustrates how these scaling laws apply to a sequence of clumps with 
constant radius and decreasing metallicity. Conversely, Figure 
\ref{clumpfig}{\em b} illustrates the scaling laws applied to a sequence 
of clumps with constant metallicity and decreasing radius.

From the geometry and scaling laws the contribution of a clump of 
radius $R$ to the number of particles $\cal N$ of each species is

\begin{eqnarray}
\cal N({\rm C}^+) &=&\left\{ 	\begin{array}{lrcl}
	{4\over3}\pi n \eta_{\rm C} (R^3-(R-\alpha h)^3) &\ \ \ \ \ \ \ \ \ \ \ \ \ \ \ \ \ \ \ \ \ \ \ \ &R&>h_+ \\
	{4\over3}\pi n \eta_{\rm C} R^3 & 0<&R&\le h_+
				\end{array}
			\right. \label{cpnp} \\
\cal N({\rm C^0}) &=&\left\{ 	\begin{array}{lrcl}
	{4\over3}\pi n \eta_{\rm C} ((R-\alpha h)^3-(R-(\alpha+1) h)^3) & &R&> h + h_+ \\
	{4\over3}\pi n \eta_{\rm C} (R-\alpha h)^3 & h_+ <&R&\le h + h_+\\
	0 &0<&R&\le h_+
				\end{array}
			\right. \label{cnp} \\
\cal N({\rm CO}) &=&\left\{	\begin{array}{lrcl}
	{4\over3}\pi n \eta_{\rm C} (R-(\alpha+1) h)^3 &\ \ \ \ \ \ \ \ \ \ \ \ \ \ \ \ \ \ \ \ \ \ \ \ \ \ &R&> h + h_+ \\
	0 &0<&R&\leq h + h_+
				\end{array} 
			\right. \label{conp}
\end{eqnarray}

\noindent where $\eta_{\rm C}$ is the carbon abundance and we are implicitly
assuming that all the carbon is in the form of the dominant 
species in each zone.
The abundance of carbon scales proportionally to $Z$, that is

\begin{equation}
\eta_{\rm C}=\eta_0 Z
\label{etascaling}
\end{equation}

\noindent where $\eta_0$ is the carbon abundance at a metallicity $Z=1$.

We will now compute the mean contribution of a clump to the column density
of each species, using CO as an example.
To eliminate the dependence on clump size, we integrate 
Equation \ref{conp} over 
the clump size distribution $R^{-\beta}$ to obtain the mean  
contribution of a clump to the number of CO molecules in the telescope beam,

\begin{equation}
\overline{\cal N({\rm CO})} = \frac{1}{\cal M} 
	\int_{R_{min}}^{R_{max}}{{\cal N(\rm CO)} R^{-\beta}\, d R} 
	\label{eco}
\end{equation}

\noindent where $R_{min}$ and $R_{max}$ are the lower and upper limits
of the clump size distribution, and $\cal M$ is the normalization integral

\begin{equation}
{\cal M} = \int_{R_{min}}^{R_{max}} R^{-\beta} \, dR 
\end{equation}

\noindent To simplify the calculations, we relate the lower and upper end 
of the clump size distribution to $h_0$ by defining $R_{min}\equiv\gamma h_0$ 
and $R_{max}\equiv\delta h_0$, where $\gamma$ and $\delta$ are unitless.

The total number of particles in the beam will be the number of clumps 
in it times the expected contribution per clump $\overline{\cal N({\rm CO})}$.
Using $\beta=3$ and the fact that $\delta\gg\gamma$ we obtain an 
algebraic expression for ${\cal N({\rm CO})}$ 
which we convert to column density (and somewhat simplify) 
by recognizing that the expectation value for the radius of a clump is 

\begin{equation}
\overline{R}=\frac{\int_{\gamma h_0}^{\delta h_0} R\ R^{-3} dR}
{\int_{\gamma h_0}^{\delta h_0} R^{-3} dR}\cong2 \gamma h_0
\end{equation}

Thus, dividing 
$\overline{\cal N({\rm CO})}$ by $\pi\overline{R}^2$ we obtain the 
mean CO column density per clump

\begin{eqnarray}
\overline{N_{\rm CO}} &=& \left\{ \begin{array}{lrcl}
	N_0
	\left[\frac{(\alpha+1)^3-6(\alpha+1)^2\delta Z+3(\alpha+1)\delta^2 Z^2
	+2\delta^3 Z^3}{2\delta^2 Z^2} \right. \\
	\left. \ \ \ \ \ \ \ \ \ \ \ 
	+3(\alpha+1)\log
	\left(\frac{1+\alpha}{\delta Z}\right)\right] & &Z&
	>\frac{\alpha+1}{\delta} \\
	0 & \frac{\alpha+1}{\delta}\geq &Z&>0
	\end{array} \right.
\label{enco}
\end{eqnarray}

\noindent where we define $N_0\equiv{2\over3} n_{\rm H}\eta_0 h_0$.
Notice that when $R_{max}\le h + h_+$ even the
biggest clump is completely photodissociated. As a result there is a critical
metallicity $Z_{\rm CO}=(1+\alpha)/\delta$ below which CO is absent from 
the ensemble of clumps.

Proceeding similarly for Eqs. \ref{cpnp} and \ref{cnp} we obtain the 
following expressions for the mean C$^+$ and C$^0$ column densities

\begin{eqnarray}
\overline{N_{\rm C^+}} &=& \left\{ \begin{array}{lrcl}
	N_0
	\left[
	\frac{-\alpha^3+6\alpha^2\delta Z-3\alpha\delta^2 Z^2}{2\delta^2 Z^2}
	+3\alpha\log\left(\frac{\delta Z}{\alpha}\right)\right] 
	\ \ \ \ \ \ \ \ & &Z&
	>\frac{\alpha}{\delta} \\
	N_0\ \delta Z
	& \frac{\alpha}{\delta}\geq &Z&>0
	\end{array} \right. \label{encp}\\
\overline{N_{\rm C^0}} &=& \left\{ \begin{array}{lrcl}
	N_0
	\left[\frac{-(1+3\alpha+3\alpha^2)+
	6(1+2\alpha)\delta Z-3\delta^2 Z^2}{2\delta^2 Z^2} \right. \\
	\left. \ \ \ \ \ \ \ \ \ \ \ 
	+3\alpha\log\left({\frac{\alpha}{\alpha+1}}\right)
	+3\log\left(\frac{\delta Z}{\alpha+1}\right)\right] & &Z&
	>\frac{\alpha+1}{\delta} \\
	N_0
	\left[\frac{\alpha^3-6\alpha^2\delta Z+3\alpha\delta^2 Z^2+
	2\delta^3 Z^3}{2\delta^2 Z^2}
	+3\alpha\log\left(\frac{\alpha}{\delta Z}\right)\right]
	&\frac{\alpha+1}{\delta}\geq &Z& >\frac{\alpha}{\delta} \\
	0 & \frac{\alpha}{\delta}\geq &Z&>0
	\end{array} \right.
\label{enci}
\end{eqnarray}

Equations \ref{enco}, \ref{encp} and \ref{enci} constitute {model A}. 

Note that in this parametrization $\alpha$, the ratio of the sizes of the
C$^+$ and C$^0$ regions, is also approximately the ratio
of the C$^+$ to C$^0$ columns times a geometrical correction factor of
order unity. This stems from the construction of the model ($\alpha$ is
the ratio between the extents of the C$^+$ and C$^0$ regions) as
well as from the limit of Equations \ref{encp} and \ref{enci}
when ${{\delta Z}\over\alpha}\gg1$ and $\alpha\sim1$ (c.f., \S\ref{parval}).
In this limit the ratio ${N_{\rm C^+}}/{N_{\rm C^0}}$ is 
approximately 

\begin{equation}
\frac{N_{\rm C^+}}{N_{\rm C^0}}\approx\alpha\frac
{3\log(\frac{\delta Z}{\alpha})-1.5}{3\log(\frac{\delta Z}{\alpha})-4.3}
\sim \alpha
\end{equation}

\noindent The value of $\delta$ determines mainly the CO column, since the
biggest clumps are mostly CO. Consequently a decrease in $\delta$ will
increase both the ${N_{\rm C^+}}/{N_{\rm CO}}$ and 
the ${N_{\rm C^0}}/{N_{\rm CO}}$ ratios. Thus the effects of 
both parameters are 
relatively well separated.

\subsection{Model B}
We will reconsider now the scaling law for the C$^0$ region, 
expressed in Equation \ref{cscaling}. If most of the C$^0$ is not originated
by photoprocesses but by other chemical reactions not strongly
dependent on UV photons, the neutral carbon
column density may not scale as $Z^{-1}$ (c.f., Equation \ref{cscaling}) 
but as a slower function 
of $Z$. As the limiting case, we will assume that the size of the C$^0$
region is constant with metallicity

\begin{equation}
h=h_0
\label{cscalingB}
\end{equation}

The size of the C$^+$ region nevertheless will continue to scale with 
$Z^{-1}$, as C$^+$ is indeed produced mainly by photoionization. 
Therefore Equation
\ref{cpscaling} still holds. The parameter $\alpha$ is then akin to
the ratio of C$^+$ to C$^0$ column densities at the relative metallicity
$Z=1$. The corresponding equations for the mean column densities of
C$^+$, C$^0$ and CO are

\begin{eqnarray}
\overline{N_{\rm C^+}} &=& \left\{ \begin{array}{lrcl}
	{N_0}
	\left[
	\frac{-\alpha^3+6\alpha^2\delta Z-3\alpha\delta^2 Z^2}{2\delta^2 Z^2}
	+3\alpha\log\left(\frac{\delta Z}{\alpha}\right)\right] 
	\ \ \ \ \ \ \ \ \ \ \ \ \ \ \
	& &Z& >\frac{\alpha}{\delta} \\
	{N_0}\ \delta Z
	& \frac{\alpha}{\delta}\geq &Z&>0
	\end{array} \right. \label{encpB}\\
\overline{N_{\rm C^0}} &=& \left\{ \begin{array}{lrcl}
	{N_0}
	\left[-\frac{3\alpha^2+3\alpha(1-4\delta) Z+(1-6\delta+3\delta^2) Z^2}
	{2\delta^2 Z} \right. \\
	\left. \ \ \ \ \ \ \ \ \ \ \ \ \ \ \ \ \
	+3\alpha\log\left(\frac{\alpha}{\alpha+Z}\right)+
	3 Z \log\left(\frac{\delta Z}{\alpha+Z}\right)\right] & &Z&
	>\frac{\alpha}{\delta-1} \\
	{N_0}
	\left[\frac{\alpha^3-6\alpha^2\delta Z+3\alpha\delta^2 Z^2+
	2\delta^3 Z^3}{2\delta^2 Z^2}
	+3\alpha\log\left(\frac{\alpha}{\delta Z}\right)\right]
	&\frac{\alpha}{\delta-1}\geq &Z& >\frac{\alpha}{\delta} \\
	0 & \frac{\alpha}{\delta}\geq &Z&>0
	\end{array} \right.\label{enciB}\\
\overline{N_{\rm CO}} &=& \left\{ \begin{array}{lrcl}
	{N_0}
	\left[\frac{\alpha^3+3\alpha^2(1-2\delta) Z+
	3\alpha(1-4\delta+\delta^2) Z^2
	+(1-6\delta+3\delta^2+2\delta^3) Z^3}{2\delta^2 Z^2} \right. \\
	\left. \ \ \ \ \ \ \ \ \ \ \ 
	+3(\alpha+Z)\log
	\left(\frac{\alpha+Z}{\delta Z}\right)\right] & &Z&
	>\frac{\alpha}{\delta-1} \\
	0 & \frac{\alpha}{\delta-1}\geq &Z&>0
	\end{array} \right.\label{encoB}
\end{eqnarray}

Equations \ref{encpB}, \ref{enciB} and \ref{encoB} constitute {model B}.
We compare the relative merits of both models in \S\ref{comparison}.
Notice that models A and B produce identical results at $Z=1$.

\section{Computing integrated line ratios}
\label{lrats}

We can now compute the line intensities using our calculated column densities.
Notice that the column density ratios between species depend only on two
non-dimensional parameters: $\alpha$ (the ratio of sizes between the C$^+$ 
and C$^0$ regions) and $\delta$ (the ratio between the sizes of the 
biggest clump and the C$^0$ region). To simplify the 
calculations we will assume optically thin emission in local thermodynamic
equilibrium (LTE) for C$^+$ and C$^0$. The line integrated intensity
(erg s$^{-1}$ cm$^{-2}$ sr$^{-1}$) is then

\begin{equation}
{\rm I} = {1\over4\pi} h \nu A_{ul}\frac{g_u e^{-{h\nu\over k T_{ex}}}}{Q(T_{ex})}\,N
\end{equation}

\noindent where $h$ and $k$ are Planck's and Boltzmann's constants;
$\nu$, A$_{ul}$ and T$_{ex}$ are respectively the transition frequency, 
Einstein A coefficient and excitation temperature; 
$g_u$ is the 
degeneracy of the upper level; $Q$ is the partition function, and $N$ is the
total column density of the species considered. Typically C$^+$ and C$^0$
have opacities $\tau\leq1$ (Zmuidzinas et al. 1988\markcite{Z88}). 
The assumption
of LTE will be valid as long as most of the emitting gas is at a density 
greater than the critical density of the transition 
($\sim2500$ cm$^{-3}$ for [C~II], 
$\sim 500$ cm$^{-3}$ for the 609 $\mu$m transition of [C~I]). 

The assumption of optically thin emission from a population in LTE is,
however, unrealistic for CO.
For computing CO (J=$1\rightarrow0$) intensities we will use the 
proportionality factor
between hydrogen column density and CO line intensity observed at Galactic 
metallicity, $X_{\rm CO}=3\times10^{20}$ 
cm$^{-2}$/(K km s$^{-1}$) (Scoville \& Sanders 1987\markcite{SS87}). 
The CO and hydrogen column densities at $Z=1$ are related via the 
CO abundance $\eta_0$, $N_{\rm CO}=N_{\rm H}\eta_0$. Thus, the integrated
line intensity (in erg s$^{-1}$ cm$^{-2}$ sr$^{-1}$) can be related to the 
CO column density in the following way

\begin{equation}
{\rm I_{CO}} = {2 k\over\lambda_{10}^3} {\overline{N_{\rm CO}}\over{X_{\rm CO}\eta_0}}
\label{ico}
\end{equation}

\noindent where $\lambda_{10}$ is the 
wavelength of the J=$1\rightarrow0$ CO transition, and the factor 
$2 k\over\lambda_{10}^3$ comes from the conversion between velocity-integrated
antenna temperature and frequency-integrated specific intensity.
In this expression
we are implicitly assuming that the emissivity of CO, determined by the 
excitation conditions of the molecule, does not depend on $Z$. 
All the dependence of
${\rm I_{CO}}$ on the metallicity stems from the metallicity dependence of 
$\overline{N_{\rm CO}}$. This assumption will be discussed further in 
\S\ref{xcosection}.

For the optically thin emission from [C II] ($^2$P$_{3/2}
\rightarrow^2$P$_{1/2}$, 158 $\mu$m) and [C I] ($^3$P$_1\rightarrow^3$P$_0$,
609 $\mu$m) we assume excitation
temperatures of 200 K and 30 K respectively (Wolfire, Hollenbach,  
\& Tielens 1989\markcite{WHT89}; Wright et al. 1991\markcite{W91}). Thus,

\begin{eqnarray}
{\rm I_{[C II]}} &=& 1.28\times10^{-21} \,\overline{N_{\rm C^+}} \nonumber \\
{\rm I_{[C I]}} &=& 7.54\times10^{-24} \,\overline{N_{\rm C^0}} \label{intensities}\\
{\rm I_{CO}} &=& 1.73\times10^{-26} \,\overline{N_{\rm CO}} \nonumber
\end{eqnarray}

\noindent where we are using the canonical value $\eta_0=3\times10^{-4}$
(e.g., Tielens \& Hollenbach 1985a\markcite{TH85a}) 
and the line intensities are in units of erg s$^{-1}$ cm$^{-2}$ sr$^{-1}$.
The equations for line ratios then are

\begin{eqnarray}
{\rm I_{[C II]}}/{\rm I_{CO}} &=& 7.4\times10^4 \,\overline{N_{\rm C^+}}/\overline{N_{\rm CO}} \nonumber \\
{\rm I_{[C I]}}/{\rm I_{CO}} &=& 4.4\times10^2 \,\overline{N_{\rm C^0}}/\overline{N_{\rm CO}} \label{ratios}\\
{\rm I_{[C II]}}/{\rm I_{[C I]}} &=& 1.7\times10^2 \,\overline{N_{\rm C^+}}/\overline{N_{\rm C^0}} \nonumber
\end{eqnarray}

\noindent where only two of the three equations are independent.

Note that because [C~II] and [C~I] are a two
and a three level system respectively, the intensities are
independent of the precise value of the temperature as long as it
exceeds the line excitation threshold.

\section{The values of the parameters $\alpha$ and $\delta$}
\label{parval}

We can use Galactic data to assign values to $\alpha$ and $\delta$.
In particular, we could use COBE's FIRAS measurements averaged over the 
entire galactic plane. Unfortunately FIRAS
did not obtain a significant detection of the CO (J=$1\rightarrow0$) 
transition. This opens a range of possibilities for the CO intensity:
{\em 1)} the formal value obtained by FIRAS 
($0.6\pm0.7$ erg s$^{-1}$ cm$^{-2}$ sr$^{-1}$; 
Wright et al. 1991\markcite{W91}),
{\em 2)} the result of a multi-line excitation fit performed with the higher
transitions of CO which FIRAS did detect (0.5 erg s$^{-1}$ cm$^{-2}$ 
sr$^{-1}$; 
Wright et al. 1991\markcite{W91}), or {\em 3)} the result of assuming 
the standard empirical ratio ${\rm I_{[C II]}/I_{CO}}=4400$ (Wolfire et al. 
1989\markcite{WHT89}), 
which corresponds to a CO (J=$1\rightarrow0$) intensity of
0.4 erg s$^{-1}$ cm$^{-2}$ sr$^{-1}$ 
given the observed FIRAS [C~II] intensity of
$1726\pm70$ erg s$^{-1}$ cm$^{-2}$ sr$^{-1}$. 
Adopting the intermediate value ${\rm I_{CO}}=0.5$ 
erg s$^{-1}$ cm$^{-2}$ sr$^{-1}$ we obtain the ratios 
${\rm I_{[C II]}/I_{[C I]}}\cong 270$
and ${\rm I_{[C I]}/I_{CO}}\cong 13$ from the FIRAS measurements. 
Numerically solving the system of Equations \ref{ratios} 
we find the values $\alpha\cong1.25$ and $\delta\cong435$. 
Note that this result is independent of the choice of model A or B,
since as was pointed out before both models are identical at $Z=1$.

\begin{figure}
\plotfiddle{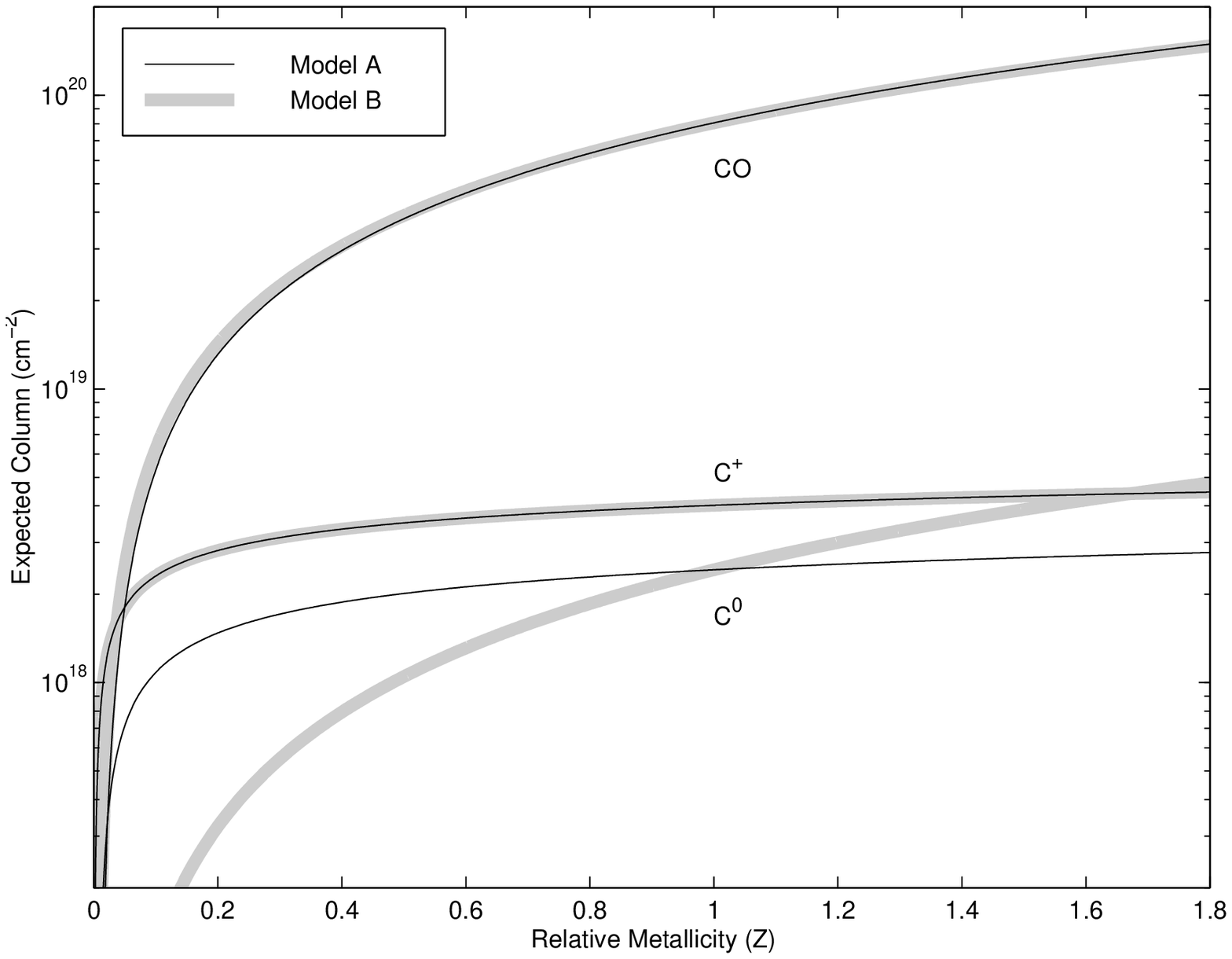}{3.0 in}{0}{70}{60}{-210}{-100}
\plotfiddle{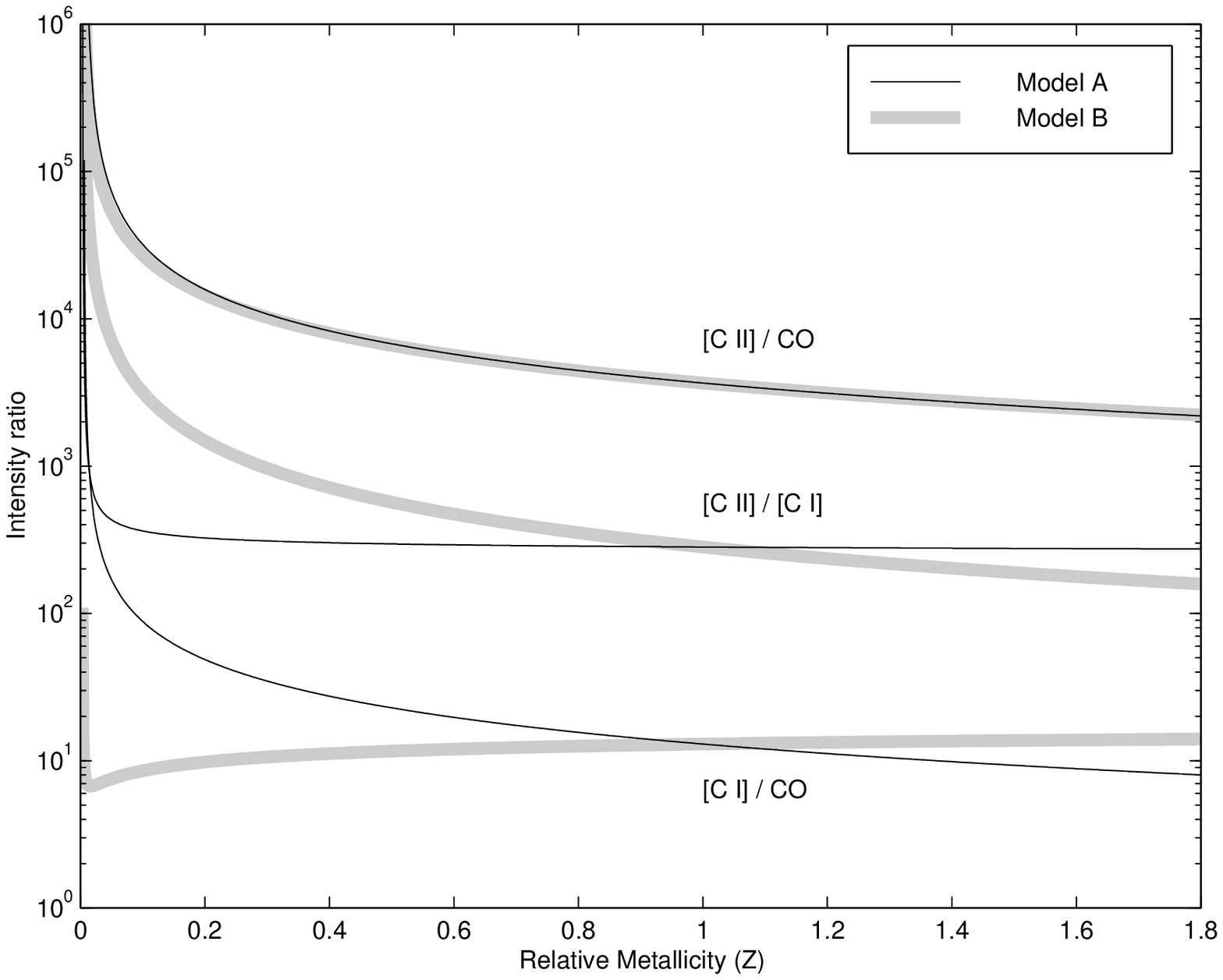}{3.0 in}{0}{70}{60}{-210}{-100}
\caption{{\em (a) Top.} The mean column densities of CO, C$^0$
and C$^+$ versus metallicity resulting from models A and B. 
The values used for the hydrogen density 
and the size of the C$^0$ region for the purpose of the plot are 
$n=10^4$ cm$^{-3}$ and $h_0=10^{17}$ cm ($\sim 0.03$ pc). 
{\em (b) Bottom.} The line intensity ratios versus metallicity for
models A and B.
Notice that, in model A, the [C II]/[C I] ratio is almost constant 
over a wide range of relative metallicities, while in model B 
the [C~I] to CO ratio is approximately constant.}
\label{intrat}
\end{figure}

The resulting column density and line intensity ratios
are shown in Figure \ref{intrat}. With the above values for 
$\alpha$ and $\delta$,
the ratio of C$^+$ to C$^0$ column densities in {model A} yields
$N_{\rm C^+}/N_{\rm C^0}\approx1.65\pm0.10$ over a wide range of 
metallicities. In {model B} the quantity that is approximately 
conserved for different metallicities is the ratio of C$^0$ to CO 
column densities, and consequently the ratio of intensities.

\begin{figure}
\plotone{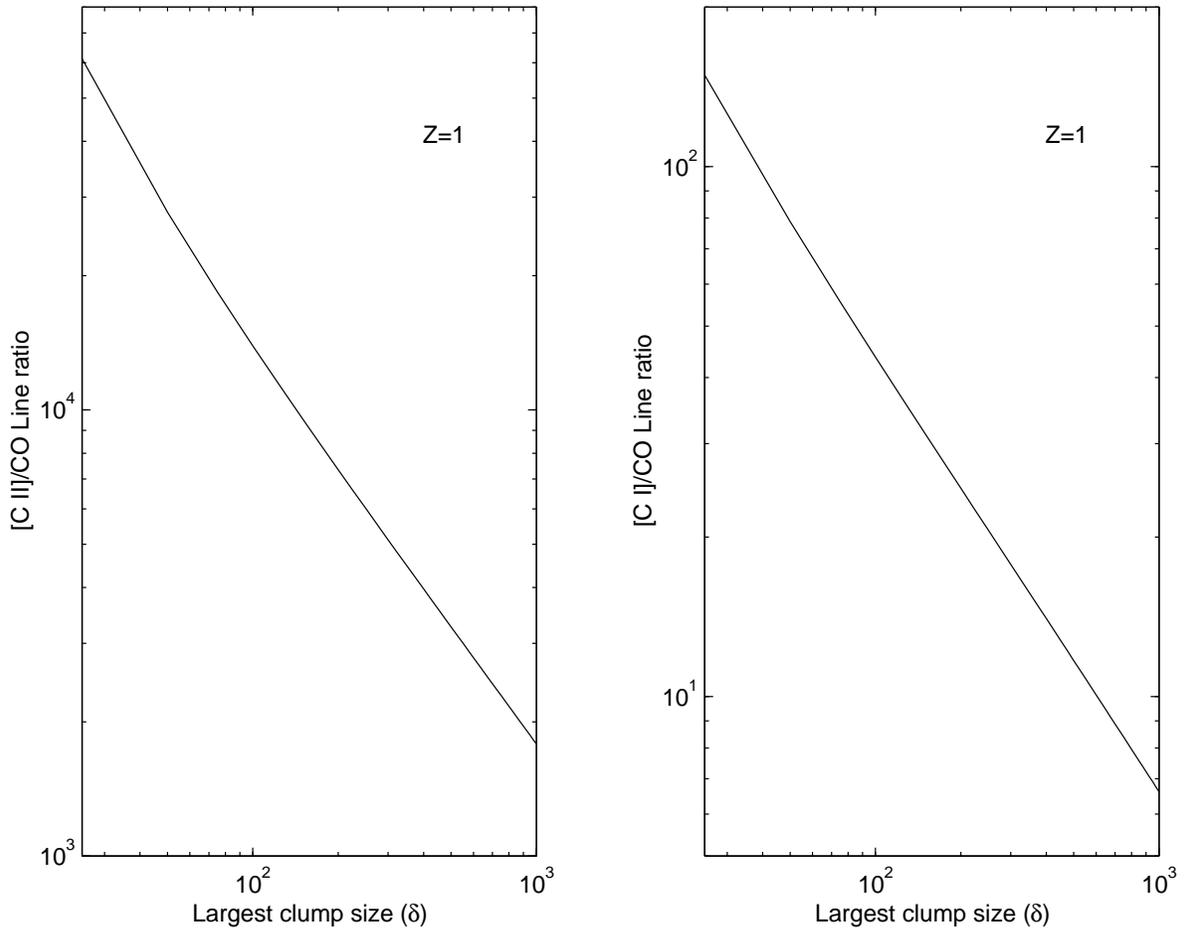}
\caption{The influence of the upper limit of the clump size distribution
(i.e., the parameter $\delta$) on the model ratios at Galactic 
metallicity ($Z=1$). The biggest clumps contribute mostly CO to the 
ensemble; consequently the line ratios decrease for higher clump sizes.}
\label{ratvsdelta}
\end{figure}

It is interesting to point out again that lowering the
value of the maximum clump size $R_{max}$ (set by $\delta$) increases 
the column densities of C$^+$ and C$^0$. This occurs because 
the biggest clumps contribute mostly molecular
(CO) gas. In fact, they dominate the CO column density. The effects on the
line ratios of varying the maximum size parameter $\delta$ are shown 
in Figure \ref{ratvsdelta}.
Some of the abnormally high [C~II]/CO intensity ratios
observed in certain sources could be due to a local change in the 
clump size distribution, caused by the disruption of the biggest clumps.

\section{The effects of density and radiation fields}
\label{densandrad}

Up to this point, our parametrization has been purely geometrical.
We have not explicitly included changes in the gas density ($n$) 
or the radiation field intensity ($\chi_{\rm uv}$).
We can introduce these changes through their effects on the parameters 
$\alpha$, $\delta$ and $h_0$.

According to chemistry calculations (van Dishoeck 
\& Black 1988\markcite{VDB88}), at a hydrogen
density $n=10^4$ cm$^{-3}$ the C$^+$ column density increases by a 
factor of $\sim 10$ when $\chi_{\rm uv}$ is
increased from 1 to $10^4$ (in units of $1.6\times10^{-3}$ erg cm$^{-2}$ 
s$^{-1}$; Habing 1968\markcite{H68}), 
while the C$^0$ column density increases roughly a
factor of $\sim 2$ to 3. Thus, the column density of C$^0$ (controlled by the
parameter $h_0$) as well as the ratio of the C$^+$ to C$^0$ column densities
(controlled by the parameter $\alpha$) are both somewhat affected by
the UV field strength, albeit not dramatically. 
The logarithmic dependence of the column densities on the radiation
field occurs because, at densities larger than $\sim10^3$ 
cm$^{-3}$, the column density of the species produced by photoprocesses varies
proportionally to the penetration depth of the UV radiation, which in
turn increases logarithmically with $\chi_{\rm uv}$ 
(Stacey et al. 1991\markcite{SG91}; see also discussion in 
Mochizuki et al. 1996\markcite{MO96}). On these grounds,
we surmise that both $\alpha$ and $h_0$ are directly proportional to
the logarithm of $\chi_{\rm uv}$. Assuming that the upper end of the 
clump size distribution, $R_{max}=\delta h_0$, remains unaffected by the
UV field strength, $\delta$ is inversely proportional to $h_0$. 

A qualitative argument can be used to understand how
changes in the density will affect $h_0$, $\alpha$ and $\delta$. 
Higher densities will yield higher recombination rates for C$^+$ and
the more active chemistry will produce more CO at the expense of C$^0$ and
C$^+$. Although it is a priori unclear how the column
densities of C$^+$ and C$^0$ depend on the density, 
chemical modelling (Ingalls 1998\markcite{I98}) shows that it is 
also approximately logarithmic.

\begin{figure}
\plotfiddle{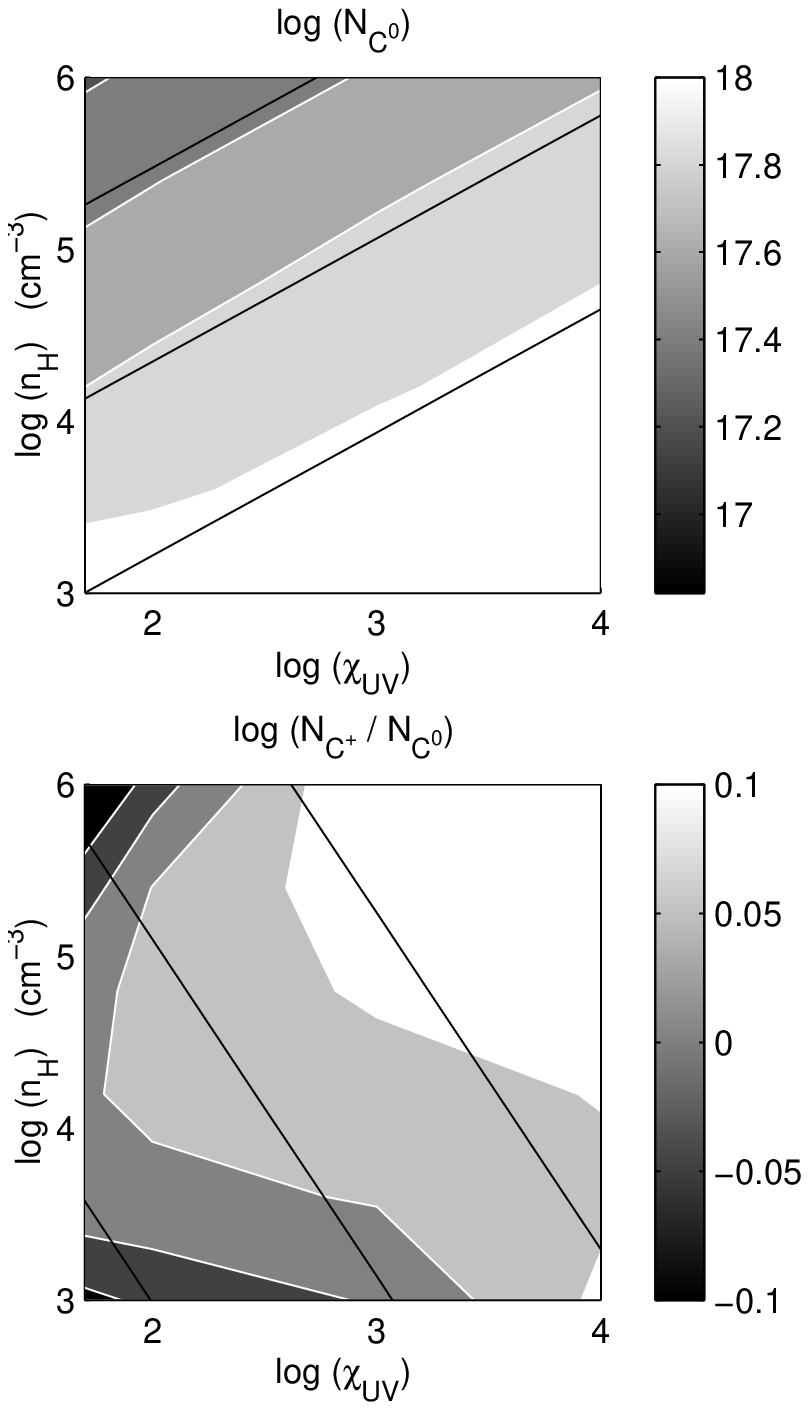}{4.5in}{0}{100}{100}{-320}{-235}
\plotfiddle{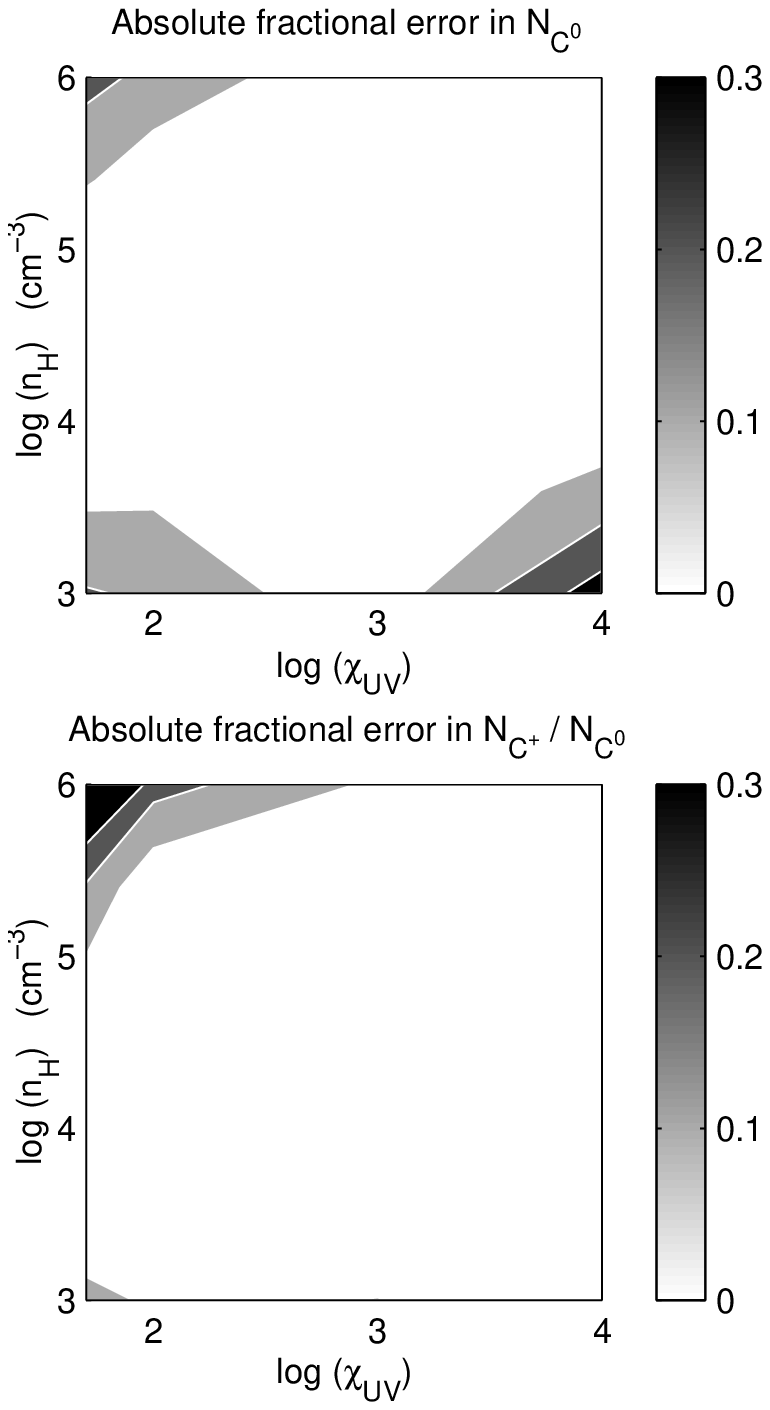}{0in}{0}{100}{100}{-320}{-204}
\caption{{\em (a) Left Column.} Two dimensional fits to the results 
of a chemical model
(Ingalls 1998\protect\markcite{I98}) for $\log N_{C^0}$ 
and $\log (N_{C^+}/N_{C^0})$. The gray scale corresponds to the results of the 
chemical network. The black contours are traced at the
same levels and show the quality of the fits. 
{\em (Top Left)} Contour levels are 10$^{17}$ 
to 10$^{18}$ cm$^{-2}$ with logarithmic spacing 0.2. 
{\em (Bottom Left)} Contour levels 
are -0.05 to 0.1 with logarithmic spacing 0.05.
{\em (b) Right Column.}
Errors introduced by using the fits ($F$) rather than the actual results of
the chemical network ($N$), expressed as $|N-F|/N$. 
The RMS error for $N_{C^0}$ and $N_{C^+}/N_{C^0}$ over the whole range of
density and UV field shown here are
11\% and 8\% respectively. Contour levels are 10\% to 30\% by 10\%.}
\label{chemfit}
\end{figure}

Figure \ref{chemfit} shows the results of the steady-state chemical network
calculations on an $A_V \geq 8$ cloud. Superimposed are the 2-dimensional
fits to the chemistry results:

\begin{eqnarray}
\log N_{C^0} &\cong& 0.13\log \chi_{\rm uv} - 0.18\log n + 18.32
\label{colcifit}\\
\log \left(\frac{N_{C^+}}{N_{C^0}}\right) &\cong& 0.05\log \chi_{\rm uv} 
+ 0.02\log n - 0.16
\label{ratfit}
\end{eqnarray}

\noindent These equations are valid over a wide range of physical conditions
($\chi_{\rm uv}\sim 50 - 10^4$; $n\sim 10^3 - 10^6$ cm$^{-3}$) and 
the errors introduced by using the fits over 
the actual chemical network results are $\approx 10 \%$.

The column densities of C$^+$ and C$^0$ obtained from Eqs. \ref{colcifit} and
\ref{ratfit} can be directly related to the parameters $\alpha$ and $h_0$ 
by observing that, to first order, $N_{C^0}\simeq n \eta_0 h_0$ and 
$N_{\rm C^+}/N_{\rm C^0}\simeq \alpha$. These two equations, combined with
$\delta\propto {h_0}^{-1}$ lead to:

\begin{eqnarray}
\alpha &\approx& 1.25\left({\chi_{\rm uv}\over{\chi_{\rm uv0}}}\right)^{+0.05} 
	\left({n\over n_0}\right)^{+0.02} \\
\label{uveffectA}
\delta &\approx& 435\left({\chi_{\rm uv}\over{\chi_{\rm uv0}}}\right)^{-0.13} 
	\left({n\over n_0}\right)^{+1.18}
\label{uveffectB}
\end{eqnarray}

\noindent where $\chi_{\rm uv0}$ and $n_0$ are the average 
radiation field and volume density of the regions that
dominate the FIRAS line emission used to assign $\alpha=1.25$ and 
$\delta=435$. Note that because HII regions probably dominate the 
FIRAS emission (Wright et al. 1991\markcite{W91}; 
Bennett et al. 1994\markcite{BE94}), 
$\chi_{\rm uv0}\gg1$. Since the FIRAS ratios are 
similar to those observed in many Galactic star-forming regions 
(in particular Orion) we normalize all values of density and
UV field to those derived for Orion, namely $n_0\sim10^4$ cm$^{-3}$ 
and $\chi_{\rm uv0}\sim250$ (Tielens \& Hollenbach 1985b\markcite{TH85b}; 
Stacey et al. 1993\markcite{S93}).

\section{Comparison with observations}
\label{comparison}

The predictions of models {A} and {B} can be directly compared with 
observational data. To properly accomplish this comparison we need 
[C~II] and [C~I] observations
on a sample of objects of the same class with different metallicities. 
Unfortunately very little [C~I] data
is available on extragalactic sources of differing metallicities (c.f., 
Wilson 1997\markcite{WI97}; Table 2) 
and only a few of them have been observed in [C~II]. 

The data we use for the comparison are listed in
Table \ref{tablesourcedata}. The selected sources are a set of star-forming
regions with different metallicities which have been observed in [C~II] and 
[C~I]. To them we have added N~27 in the Small Magellanic Cloud (SMC), 
a source that 
does not yet have [C~I] observations but which we expect will be important 
to probe our models because it has the lowest metallicity in the sample.

\begin{figure}
\plotone{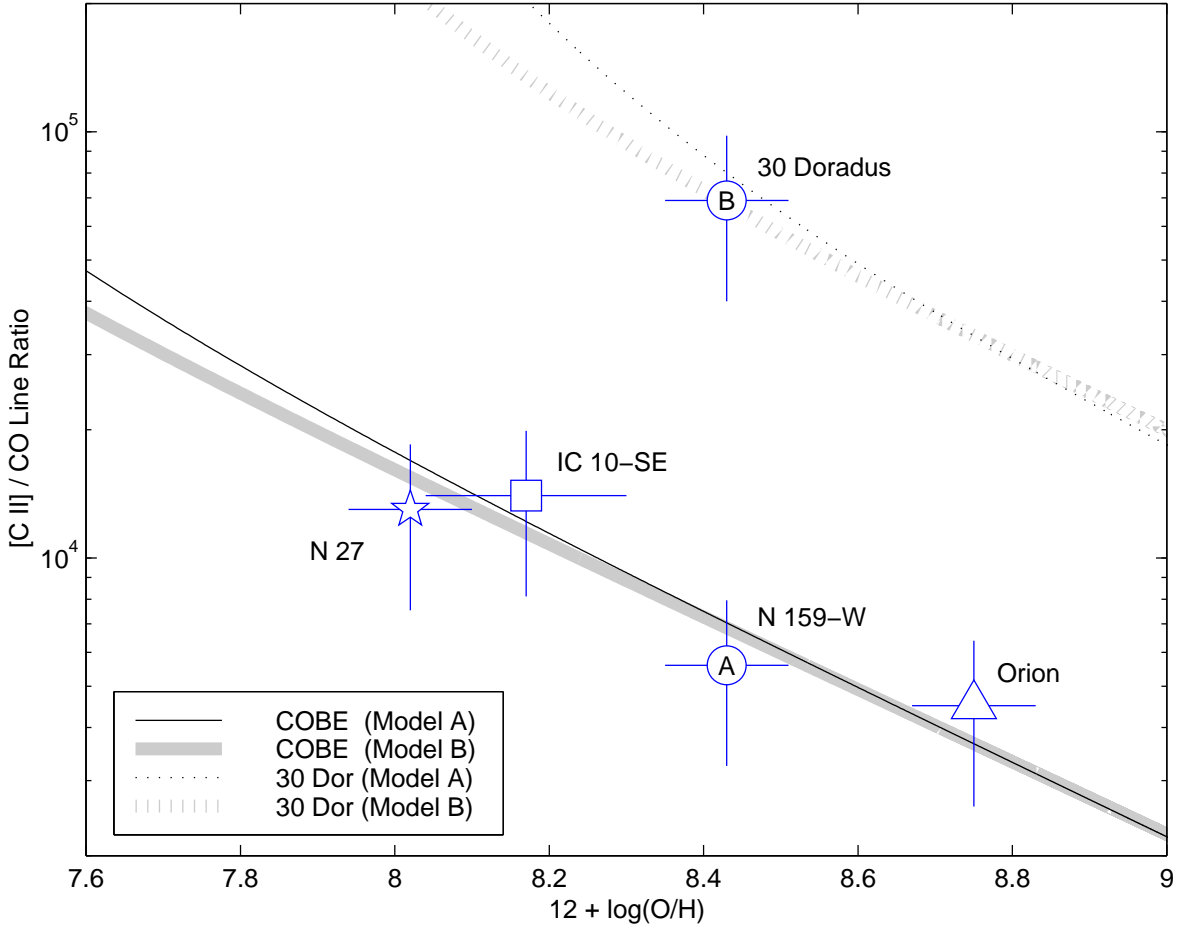}
\caption{Comparison of model A and B predictions with [C II]/CO
intensity ratios observed in several sources. 
The solid lines are the calculated [C II]/CO (J=$1\rightarrow0$) 
intensity ratio with $\alpha$ and $\delta$ derived from the COBE FIRAS ratios. 
The dotted line corresponds to the COBE-based $\alpha$ and $\delta$ corrected 
in the way prescribed in \S\protect\ref{densandrad} by
assuming that for 30 Doradus $\chi_{\rm uv}=3500$ and $n=2000$ cm$^{-3}$.
The data points are: the Orion interface region (triangle), 
N 159 in the LMC (circle A), 30 Doradus (circle B), 
IC 10 SE (square), and N 27 in the SMC (star). 
The error bars in the ratio are calculated 
assuming 30 \% calibration accuracy in both lines.}
\label{cp2corat}
\end{figure}

Figure \ref{cp2corat} compares the I$_{\rm [C II]}$/I$_{\rm CO}$ intensity 
ratio predictions of models {A} and {B} with the observational
data, using the values for $\alpha$ and $\delta$ derived from 
FIRAS measurements. Both models give essentially identical predictions for the
decreasing trend in the I$_{\rm [C II]}$/I$_{\rm CO}$ intensity ratio 
with increasing metallicity. This similarity between models {A} and {B} 
should not be surprising, since the effect of keeping $h$ constant in 
a clump (as model B does) is to enlarge its CO region by only a 
small fraction for $Z<1$, with respect to the same clump in model {A}. 
The fact that the 
I$_{\rm [C II]}$/I$_{\rm CO}$ ratio for most sources (except 30 Doradus) 
falls along the model line can be interpreted as showing that N~27,
IC~10-SE, N~159 and Orion all share similar 
values of $n$ and $\chi_{\rm uv}$. Another possibility is that the density and 
UV field in these regions conspire in such a way as to keep their line
ratios close to the model line. This can happen because, as shown
in Equation \ref{uveffectB}, the effects of density and radiation
field go in opposite directions and have the potential to 
cancel each other. Denser regions, which are expected to have lower column
densities of C$^+$ and C$^0$,  will be in general more efficient at 
producing stars and thus will have on average a higher UV field which
in turn will increase $N_{\rm C^+}$ and $N_{\rm C^0}$. 
However, because of the large difference in the exponents in Equation 
\ref{uveffectB}, this cancellation is unlikely to occur and we can safely
assume that these star-forming regions posses similar densities and radiation
fields.

The large ${\rm I_{[C II]}/I_{CO}}$ ratio observed 
in 30 Doradus suggests either 
that the gas in this source is under quite different physical conditions,
or that $R_{max}$, the upper limit of the clump size distribution, is actually
smaller than for the rest of the sample. Lowering the size of the 
biggest clump (by making $\delta$ smaller) will produce more 
photoionized material at the expense of CO, which will in turn raise the
observed line ratio (c.f., \S\ref{parval}). The gas in 30 Doradus is under
an extremely intense (and hard) UV field, $\chi_{\rm uv}\simeq3500$ 
(Poglitsch et al. 1995\markcite{PO95}). From the numerical exponents
of the relationships postulated in \S\ref{densandrad} (c.f., Equations
\ref{uveffectA} and \ref{uveffectB}), an
increase by a factor of 15 in $\chi_{\rm uv}$ with respect to Orion 
(c.f., Table \ref{tablesourcedata}) is not enough to explain the observed 
line ratio. The average gas density must also decrease to about 
2000 cm$^{-3}$, about 5 times lower than Orion's.
This density is within the range allowed by previous determinations
(Poglitsch et al. 1995\markcite{PO95}). 
Thus, it is not necessary to invoke a smaller
maximum clump size to explain the observed [C~II]/CO 
line ratio, although it remains a logical 
possibility (e.g., Pak et al. 1998\markcite{PA98}). 
The 30 Doradus nebula is an especially 
active massive star-forming region, and as such it is the site of several 
violent events which could disrupt the larger molecular clumps.

\begin{figure}
\plotone{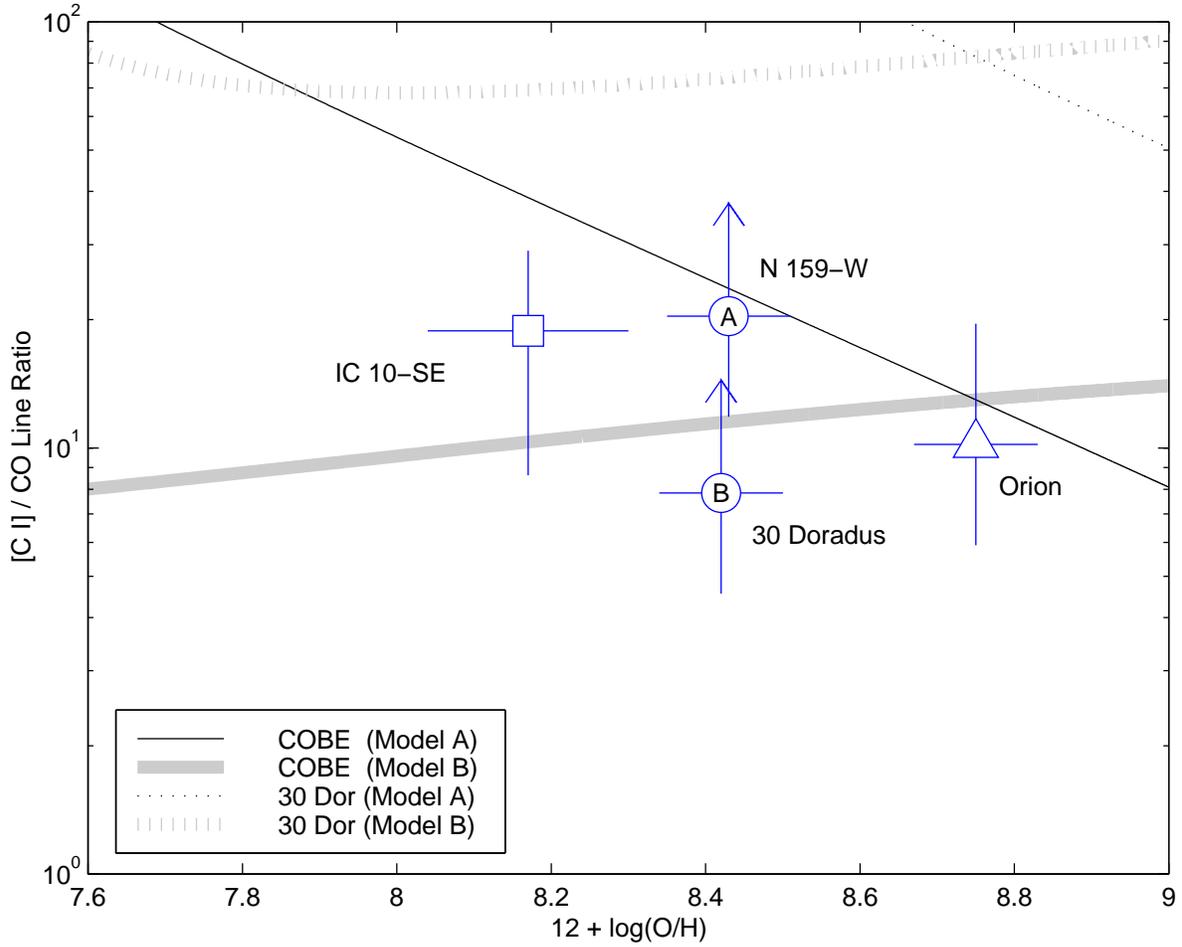}
\caption{Same as figure \protect\ref{cp2corat} but for [C~I]/CO
intensity ratios versus metallicity. The point for IC~10-SE correspond
to the average ratio in the complex 
(Bolatto et al. 1998\protect\markcite{BO98}). The point for 30 Doradus
has been slightly displaced in metallicity to eliminate confusion in the
error bars.}
\label{ci2corat}
\end{figure}

Figure \ref{ci2corat} illustrates the comparison between 
the I$_{\rm [C I]}$/I$_{\rm CO}$ intensity ratio predictions of models 
{A} and {B} and the observations. In this case the predictions
of the two models differ drastically. While {model A} predicts a ratio
that rapidly increases for decreasing metallicities, such a trend is not
apparent in the observational data, which tends to cluster 
around a ratio of $\sim10$, essentially the prediction of {model B}.
Both Large Magellanic Cloud (LMC) points obtained by AST/RO (Stark et al. 
1997\markcite{SB97}) are lower limits to the 
ratio because of pointing uncertainties. 
They are roughly consistent with either model at the metallicity of the LMC, 
although N~159 favors {model A} while the discrepancy in the case of 30 
Doradus is smaller with {model B}. A stronger constraint 
is provided by the IC~10-SE [C~I] observations 
(Bolatto et al. 1998\markcite{BO98}), which
would have to be a factor of $\sim2$ more intense in [C~I] in order 
to be compatible with {model A}. An added complexity in the interpretation
of the ratios in IC~10-SE is that the [C~II] observations have a much
larger beam size than the [C~I] (approximately 60 arcsec and 10 arcsec 
respectively). Therefore, the I$_{\rm [C II]}$/I$_{\rm CO}$ ratio may
be sampling parcels of gas under much different physical conditions from
those sampled by the I$_{\rm [C I]}$/I$_{\rm CO}$, a fact that could
make the ratio observed in IC~10-SE appear artificially low.

Stronger constraints on the models would be provided by [C~I]
observations of the SMC. If this trend of constant ${\rm [C~I]/CO}$ intensity
ratio is confirmed, we are led to conclude that most of the C$^0$ in
the clumps is not created by photodissociation but by other chemical processes.
Otherwise, we would expect the trends of [C~II]/CO and [C~I]/CO intensity
ratios with metallicity to be similar, as in model A.

\section{The value of $X_{\rm CO}$}
\label{xcosection}

The proportionality between the CO emission intensity and the column density 
of H$_2$, $N_{\rm H_2}=X_{\rm CO}$I$_{\rm CO}$, has been observationally 
established using different methods (see Magnani \& Onello 1995 
\markcite{MO95} for a review of the methods and their caveats). 
The proportionality is well accepted for
giant molecular clouds (GMCs), although the precise mechanisms by which
$X_{\rm CO}$ becomes relatively independent of the physical conditions
in the cloud remain unclear (e.g., McKee 1989\markcite{MK89}). 

There are a few observational determinations in the literature of the 
dependence of $X_{\rm CO}$ on metallicity $Z$. Using 
interferometric CO observations to obtain virial mass estimates of 
extragalactic GMC complexes, Wilson (1995)\markcite{WI95} 
found a power law dependence with a slope of $-0.67$. 
Using a combination of single-dish and 
interferometric CO data and virial mass estimates, 
Arimoto et al. (1996)\markcite{AR96} find a slope of $-0.8$.

In the context of the models presented here, we can use Equation \ref{ico}
to investigate the expected dependence of $X_{\rm CO}$ on metallicity.
Accordingly

\begin{equation}
\frac{X_{\rm CO}(Z)}{X_{\rm CO}(Z=1)}=
\frac{\overline{N_{\rm CO}}(Z=1)}{\overline{N_{\rm CO}}(Z)}
\label{xco}
\end{equation}

\noindent and thus $X_{\rm CO}$ is inversely proportional to the mean
column density of CO. The resulting slope with metallicity, according to 
Equations \ref{enco} or \ref{encoB}, is $\approx-1$.

\begin{figure}
\plotone{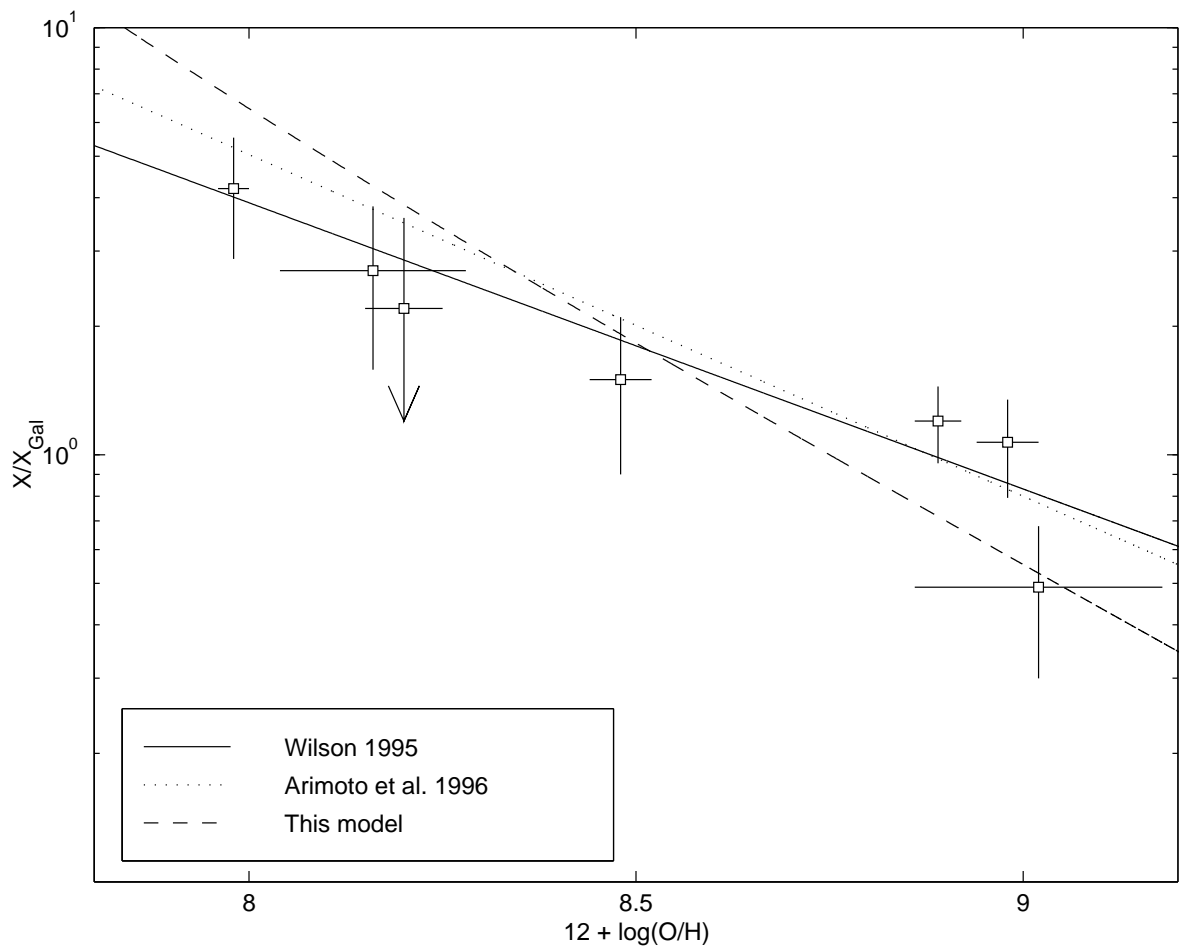}
\caption{The CO (J=$1\rightarrow0$) intensity to hydrogen column density
conversion factor ($X_{CO}$) versus metallicity, as predicted by the models
and deduced from observations. The data points are those used by 
Wilson (1995)\protect\markcite{WI95}.
The solid line corresponds to Wilson's best fit with a slope of $-0.67$. The
dotted line is a similar result obtained by 
Arimoto et al. (1996)\protect\markcite{AR96} using an
independent data set. The slope of Arimoto's best fit is $-0.8$. 
The dashed line represents the result of the models here developed, 
as explained elsewhere in the text. The slope is $-1.05$. 
Models A and B produce nearly identical results.}
\label{xfactor}
\end{figure}

A comparison of our models with the results from Wilson (1995)\markcite{WI95} 
and Arimoto et al. (1996)\markcite{AR96} is shown in 
Fig. \ref{xfactor}. The precise value of the clump size distribution 
exponent $\beta$ or the parameters $\alpha$ and $\delta$ do not affect
the value of this slope, which mainly depends on the dominant term of the 
polynomial part of $\overline{N_{\rm CO}}$. Although the Wilson 
(1995)\markcite{WI95} data 
points certainly fit a $-0.67$ slope better, the quality of the data does not
exclude a slope closer to $-1$. A similar situation occurs with the Arimoto et
al. (1996)\markcite{AR96} data. 

One of the reasons why the actual slope could be different from the 
theoretical expectation is a systematic change in the CO excitation
temperature with metallicity. Such a dependence might be expected, since
both the heating ($\Gamma$) and cooling ($\Lambda$) of the gas are 
affected by metallicity.
The $\Gamma$ term is fundamentally controlled by photoelectric ejection
of electrons from dust grains, which collisionally heat the gas. The
cooling term is controlled by emission in the [C~II], [O~I] 
and [C~I] fine structure transitions. The balance of the system
at different metallicities and dust-to-gas ratios has been computed 
by Wolfire et al. (1995)\markcite{WO95}, who conclude that
the temperature of the gas is higher for low $Z$. If the excitation
temperature of CO is higher at low metallicities, the column of
CO needed to produce a given intensity I$_{\rm CO}$ will drop, together 
with the corresponding column density of H$_2$. Consequently, there
will be a drop in the $X_{\rm CO}$ 
factor as $X_{\rm CO}=N_{\rm H_2}/{\rm I_{\rm CO}}$
which will tend to flatten the curve of $X_{\rm CO}(Z)$.

\section{Summary and conclusions}

We have presented a model of PDRs of varying metallicity
that depends on only two free parameters ($\alpha$, the ratio of 
sizes between the C$^+$ and C$^0$ regions and $\delta$, a measure
of the size of the biggest clump in the ensemble). This approach
naturally incorporates the clumpiness of the interstellar medium into 
the calculations and, when combined with some simple scaling laws,
allows us to study the influence of metallicity on the PDR line emission. 
We have shown how to modify it to include the effects of density and 
radiation field, and we have used it to study the dependence of the 
${\rm I_{[C II]}/I_{CO}}$ and ${\rm I_{[C I]}/I_{CO}}$ line ratios on the 
metallicity. In comparing our calculations with observational data, we 
find that the model explains well the observed trend of enhanced 
${\rm [C~II]/CO}$ line ratio with decreasing metallicity as the natural result
of the augmented fraction of photodissociated gas in a clump. We have shown
that enhanced ratios can be produced by lowering the density or increasing
the radiation field in the ISM, but also by lowering the upper limit
of the clump size distribution as may happen in 30 Doradus 
(Pak et al. 1998\markcite{PA98}).
We find that the available data favors a ${\rm [C~I]/CO}$ intensity 
ratio essentially independent of metallicity, albeit the paucity of 
observations does not exclude other possibilities. 
This is difficult to understand if most of the C$^0$ is produced by 
UV photons as is the case with C$^+$, arguing for a chemical
origin for at least part of the neutral carbon in these star-forming regions.
More [C~II] and [C~I] observations on low metallicity sources such as the 
Small Magellanic Cloud are necessary, however, to establish this point.
Finally, we have studied the prediction of the model for the trend of
the $X_{\rm CO}$ factor with metallicity and compared it to previous 
observational studies, and we find that the agreement is reasonable.

\acknowledgements
We thank D. P. Clemens and T. M. Bania for helpful comments on the draft 
of this manuscript. This research has made use
of NASA's Astrophysics Data System Bibliographic Services. 
This research was supported in part by the National
Science Foundation under a cooperative agreement with the Center for 
Astrophysical Research in Antarctica (CARA), grant number NSF OPP 89-20223.
CARA is a National Science Foundation Science and Technology Center.

\newpage
\begin{deluxetable}{lcccrcl}
\tablewidth{0pc}
\scriptsize
\tablecaption{Source metallicities and intensity ratios}
\tablehead{
Source& 12+log(O/H)& $\chi_{\rm uv}$ & $n$ & {I$_{\rm [C II]}$/I$_{\rm CO}$}$^a$& {I$_{\rm [C I]}$/I$_{\rm CO}$}$^a$& References\\
      &            &             & (cm$^{-3}$) \\
}
\startdata
Milky Way	  &(8.75)$\ ^b$& & & 3450$\phn^c$ 
		& \phn\phd13$\ ^c$ & Wright et al. 1991, Bennet et al. 1994\\
Orion		  &8.75\phm{$\ ^b$}& 250 & $10^4$\phm{$\ ^d$} & 4500\phn\phn 
		& 10 - 15 & Lequeux et al. 1979\markcite{LE79}, 
			    Stacey et al. 1993\markcite{S93},\\
					&&&&&& Tauber et al. 1995\markcite{TA95} \\
N 159 (LMC)	  &8.43\phm{$\ ^b$}& 300 & $8\times10^3\ ^d$ & 5600\phn\phn 
		& $\geq20$ & Dufour 1984\markcite{DU84}, 
			     Israel et al. 1996\markcite{IS96}, \\
					&&&&&& Stark et al. 1997\markcite{SB97}\\
30 Doradus (LMC)  &8.43\phm{$\ ^b$}& 3500& $10^3$ - $10^4$\phm{$\ ^d$}& 69000\phn\phn 
		& $\geq8$ & Poglitsch et al. 1995\markcite{PO95}, 
			    Stark et al. 1997\markcite{SB97}\\
IC 10 SE (IC 10)  &8.17\phm{$\ ^b$}& $\ldots$& $\ldots$& 14000\phn\phn 
		& \phn\phd19$\ ^e$ & Lequeux et al. 1979\markcite{LE79}, \\
					&&&&&& Madden et al. 1997\markcite{MA97}, Bolatto et al. 1998\markcite{BO98}\\
N 27 (SMC)	  &8.02\phm{$\ ^b$}& $\ldots$&$\ldots$ & 13000\phn\phn 
		& \ldots & Dufour 1984\markcite{DU84}, Israel et al. 1993\markcite{IS93} \\
					&&&&&& Rubio et al. 1993\markcite{RLB93} \\
\tablenotetext{a}{Intensities in erg cm$^{-2}$ s$^{-1}$.}
\tablenotetext{b}{Assumed to be identical to Orion's.}
\tablenotetext{c}{Computed using I$_{\rm CO}$=0.5 erg cm$^{-2}$ s$^{-1}$.}
\tablenotetext{d}{Also from multiline excitation study using Johansson 
et al. (1994)\markcite{JO94} observations.}
\tablenotetext{e}{Error-weighted average for points SL2, MC1, MC2 and MC3.}
\enddata
\label{tablesourcedata}
\end{deluxetable}

\newpage

\newpage

\newpage

\newpage

\newpage

\newpage

\newpage
\end{document}